\documentclass[preprintnumbers,prb,twocolumn,10pt]{revtex4}%
\usepackage{graphicx}
\usepackage{dcolumn}
\usepackage{bm}
\usepackage{latexsym,amsthm,amssymb}
\usepackage{amsmath}
\usepackage{amsfonts}
\usepackage{amssymb}%

\begin{document}
\title{The spiral spin state in a zigzag spin chain system}
\author{Meihua Chen}
\author{Chong Der Hu }
\email{cdhu@phys.ntu.edu.tw}
\affiliation{Physics Department, National Taiwan University, Taipei, R. O. C. }
\date{\today }

\begin{abstract}
We considered a spin chain with nearest neighbor and next nearest neighbor
exchange interactions, anisotropic exchange interaction and
Dzyaloshinskii-Moriya interaction. The conditions of the spiral spin state as
the ground state were analyzed. Our purpose is to build the connection between
the spiral state and the fully polarized state with a unitary transformation.
Under this transformation, anisotropic exchange interaction and
Dzyaloshinskii-Moriya interaction can be transformed to each other. Then we
use positive semi-definite matrix theorem to identify the region of fully
polarized state as the ground state for the transformed Hamiltonian, and it is
the region of spiral spin state as the ground state of the original
Hamiltonian. We also found that the effect of Dzyaloshinskii-Moriya
interaction is important. Its strength is related to the pitch angle of spiral
spins. Our method can be applied to coupled spin chains and two dimensional
triangular lattice systems. Our results can be compared with the experiment data.

\end{abstract}
\maketitle

\section{Introduction}

The spiral spin states has been the subject of study for more than sixty years
\cite{AY59,TN62,TN59}. Yet it still gives rise to surprising physical
properties. We focus on the multiferroic phenomenon found recently in numerous
compounds \cite{Wan09}. Experiments \cite{MK05,Che08} showed that in the
multiferroic material the magnetic and ferroelectric orders are closely
related. What is more intriguing is that only certain types of magnetic
orders, can couple to ferroelectricity. Furthermore, only spiral spins
configuration gives rise to strong coupling between electric polarization and
magnetic order. The spin-current model\cite{HK05} provided a plausible
explanation for this phenomenon.

Beside multiferroics, the spiral spin state was found in many other transition
metal compounds. In the early study\cite{Kim08,TS09}, they have presented the
evidences of the existence of spiral spin state in multiferroics. In
LiCu$_{2}$O$_{2}$ \cite{neutron,li08,li09} and NaCu$_{2}$O$_{2}$ \cite{Na10}, there
are one-dimensional spin chains consist of edge-sharing octahedra. The Cu-O-Cu
bond angle is almost $90$-degree. This renders the superexchange interaction
between nearest neighbor (NN) weak and ferromagnetic \cite{Khom96} and the
exchange between next nearest neighbor (NNN) not negligible. The spiral spin
configuration can also be found in higher dimension systems. The structure of
ACrO$_{2}$ (A=Cu, Ag, Li, or Na)\cite{To08} is a two dimensional triangular
lattice, and its bond angle of Cu-O-Cu is also close to $90$-degree. This can
also be the cause of the spiral spin state.

The existence of spiral spin configuration can be attributed to the
frustration in the system. A relatively simple case is a one-dimensional spin
chain with NN and NNN exchange interactions, to be denoted as $J_{1}$\ and
$J_{2}$ respectively. Frustration is caused by their competing tendencies of
aligning spins. This kind of systems is often called zigzag spin chains. There
have been much analysis on this subject. Exact solutions have been found for
special cases. The most notable case is the dimer state at $J_{1}/J_{2}=2$
found by Majumdar and Ghosh \cite{MG69}. At the other end $J_{1}/J_{2}=-4$, it
was found that the fully polarized (FP) state and uniformly distributed
resonant valence bond state \cite{Nat88} are degenerate ground states (GS). In
general, the phase diagram is summarized by Bursill \cite{RB95}. The boundary
of the frustrated region is identified by numerical calculation. White and
Affleck \cite{White96} calculated the correlation function $\left\langle
\overrightarrow{S}_{i}\overrightarrow{S}_{j}\right\rangle $ and provided solid
evidence of the existence of the spiral spin state. Furthermore, it has been
found that in zigzag spin chains, there exists chiral order
\cite{TH10,IP08,KO08}. This is another indication of the extensive existence
of spiral spin state. Hence the existence of spiral spin state becomes an
important subject.

The spiral spin state can be found in many physical systems.
For example, both of neutron diffraction\cite{neutron} and polarization dependent
resonant soft x-ray magnetic scattering (RSXMS)\cite{li08} experiments
indicate an incommensurate superstructure with $Q=(0.5,0.1738,0)$  at low temperature,
where $\vec{Q}$ is the wave vector of spiral spins.
In the direction of chain (b-axis) the spiral angle is about $\phi'=\vec{Q}\cdot\vec{b}=62^0$,
which is the angle difference of two adjoint spin.
We will show our result in the case of $\phi'=50^0$ which is comparable to experimental finding.
The discrepancy could come from inter-chain coupling.
Though we also consider the inter-chain coupling in Sec.\ref{extension},
its magnitude may not be practical. Further work is needed.

Dzyaloshinskii-Moriya (DM) interaction \cite{DM58,DM60} not only plays an
important role in the multiferroic material, but also produces many exotic
physical phenomena. It is another mechanism which can give rise to spiral spin
state. It can act as a vector potential on the spin wave in the magnon spin
Hall effect \cite{Hall10}. In ferromagnetic nanowires DM interaction has
profound effect on the motion of domain walls \cite{AA10}. It can also give
rise to spin current and soliton in spin chains\cite{GJA08}. Therefore, it is
important to incorporate DM interaction into the model Hamiltonian to see what
role it plays.

The purpose of our study is to find the condition for the spiral spin state
being the ground state (GS) in zigzag spin chain. In Sec.\ref{2connect}, we
start with a Hamiltonian with NN, NN and DM interactions and derive the
conditions of spiral spin states being the eigen states by performing a
unitary transformation. In Sec.\ref{localsite}, we use the positive
semi-definite (PSD) matrix theorem to determine the conditions of ground state
(GS), by decomposing the system into local Hamiltonian. Examples are given in
Sec.\ref{iso-h} and Sec.\ref{nodm} to illustrate our result. The problem of
symmetry is discussed in Sec.\ref{symmetry}. The applications to real physical
systems of coupled zigzag spin chain and two-dimensional triangular lattice
are given in Sec.\ref{extension}. In Sec. \ref{compare}, we compare our
results with those of numerical calculations and simulations. Sec.
\ref{summary} is devoted to the conclusions.


\section{Spiral spin state as an eigen state\label{2connect}}

The physical systems which prefer spiral spin configuration usually have
competing interactions. For example, when the nearest neighbor (NN) exchange
interaction is weak, the next nearest neighbor (NNN) interaction or even
Dzyaloshinskii-Moriya (DM) interaction become relatively significant. The
Hamiltonian of this kind of systems (are usually called the zigzag spin
chains) can be written as
\begin{align}
H=  &  \sum_{j}J_{1} [\Delta_{1}s_{j}^{z}s_{j+1}^{z}+\frac{1}{2}(s_{j}%
^{+}s_{j+1}^{-}+s_{j}^{-}s_{j+1}^{+})\nonumber\\
&  +D_{1}\vec{s}_{j}\times\vec{s}_{j+1}\cdot\hat{z}]\nonumber\\
&  +J_{2}[\Delta_{2}s_{i}^{z}s_{j+2}^{z}+\frac{1}{2}(s_{j}^{+}s_{j+2}%
^{-}+s_{j}^{-}s_{j+2}^{+})\nonumber\\
&  +D_{2}\vec{s}_{j}\times\vec{s}_{j+2}\cdot\hat{z}] \label{eq-H}%
\end{align}
where $j$ is the label of the lattice site, $J_{1}$ is the NN interaction,
$J_{2}$ is the NNN interaction, $\Delta_{1}(\Delta_{2})$ gives the anisotropic
interaction along $z$ axis for NN (NNN) interaction, and $D_{1}(D_{2})$ is the
DM interaction between NN (NNN).

\begin{figure}[h]
\begin{center}
\includegraphics[scale=0.45]{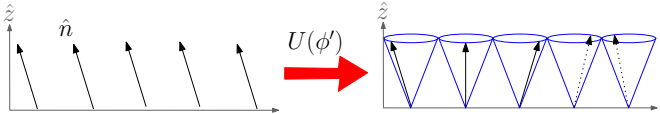}
\end{center}
\caption{The way to analyze spiral state is to connect it to a fully polarized
(FP) state with a unitary transformation, which is the product of rotations
around $z$-axis. }%
\label{fp-spiral}%
\end{figure}
Although there have been numerous studies on the zigzag spin chain, the
boundary for the spiral spin state being the ground state still cannot be
determined if DM interaction is present. Here we propose another way to
analyze. We connect the spiral spin state and a fully polarized (FP) state by
a unitary transformation. The FP state and the spiral spin state are shown in
Fig. \ref{fp-spiral}. We then ask the question: Under what conditions the FP
state will be the ground state (GS) of the transformed Hamiltonian? Since
under any unitary transformation, the energy spectrum does not change, the
ground states of the physical Hamiltonian and transformed Hamiltonian are
equivalent. By identifying the region of FP state being GS for the transformed
Hamiltonian, we will know the region of spiral spin state as GS of the
physical Hamiltonian.
\begin{figure}[t]
\begin{center}
\includegraphics[scale=0.5]{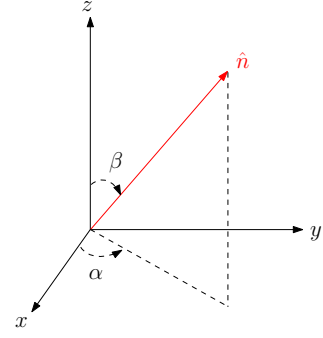}
\end{center}
\caption{The definition of spin orientation $\hat{n}$, where $\hat{n}%
=(\sin(\beta)\cos(\alpha),\sin(\beta)\sin(\alpha),\cos(\beta))$.}%
\label{direction-spin}%
\end{figure}

The unitary transformation mentioned above rotates the spins around the
$z$-axis\cite{LOA92}. It has the form
\begin{equation}
U(\phi^{\prime})=\prod_{j=1}^{N}\exp(is_{j}^{z}\vec{Q}\cdot\vec{R}_{j}),
\end{equation}
with $\phi^{\prime}=\vec{Q}\cdot(\vec{R}_{j+1}-\vec{R}_{j})$ being a constant.
For now, $\phi'$ and hence, $\vec{Q}$ can have arbitrary values.
But, as it will be shown later, the DM interaction has a profound effect on $\phi'$ and $\vec{Q}$.
For a given strength of DM interaction ($D_1$ and $D_2$), the boundaries of $\phi'$ and $\vec{Q}$ are determined.
Within the boundaries, the spiral spin states are the ground states of the system.
Hence, the spiral spin state
\begin{equation}
|\psi_{ss}\rangle=\prod_{j=1}^{N}[\cos(\frac{\beta}{2})|\uparrow\rangle
_{j}+\sin(\frac{\beta}{2})e^{i(\vec{Q}\cdot\vec{R}_{j}-\alpha)}|\downarrow
\rangle_{j}],\nonumber
\end{equation}
can be transformed into a FP state with spin direction $\hat{n}=(\sin
(\beta)\cos(\alpha),\sin(\beta)\sin(\alpha),\cos(\beta))$ (see Fig.
\ref{direction-spin}), as the following form
\begin{equation}
|FP,\hat{n}\rangle=\prod_{j=1}^{N}[\cos(\frac{\beta}{2})|\uparrow\rangle
_{j}+\sin(\frac{\beta}{2})e^{-i\alpha}|\downarrow\rangle_{j}].
\end{equation}
The Hamiltonian is also transformed
\begin{align}
&  H_{rot}=UHU^{-1}\nonumber\\
=\sum_{j} &  [J_{1}(\Delta_{1}s_{j}^{z}s_{j+1}^{z}+\frac{\cos(\phi_{1}%
-\phi^{\prime})}{2\cos(\phi_{1})}(s_{j}^{+}s_{j+1}^{-}+s_{j}^{-}s_{j+1}%
^{+})\nonumber\\
&  +\frac{\sin(\phi_{1}-\phi^{\prime})}{\cos(\phi_{1})}\vec{s}_{j}\times
\vec{s}_{j+1}\cdot\hat{z})\nonumber\\
&  +J_{2}(\Delta_{2}s_{j}^{z}s_{j+1}^{z}+\frac{\cos(\phi_{2}-2\phi^{\prime}%
)}{2\cos(\phi_{2})}(s_{j}^{+}s_{j+2}^{-}+s_{j}^{-}s_{j+2}^{+})\nonumber\\
&  +\frac{\sin(\phi_{2}-2\phi^{\prime})}{\cos(\phi_{2})}\vec{s}_{j}\times
\vec{s}_{j+2}\cdot\hat{z})]\nonumber
\end{align}
where we have set $\phi_{1}=\tan^{-1}(D_{1}),\phi_{2}=\tan^{-1}(D_{2})$, and
used the identity $1+i\tan(\phi)=\sec(\phi)e^{i\phi}$.

In order that $|FP,\hat{n}\rangle$ is an eigen-state of the transformed
Hamiltonian, it is required that
\begin{align}
\Delta_{1} &  =\frac{\cos(\phi_{1}-\phi^{\prime})}{\cos(\phi_{1}%
)},\label{delta1}\\
\Delta_{2} &  =\frac{\cos(\phi_{2}-2\phi^{\prime})}{\cos(\phi_{2}%
)}.\label{delta2}%
\end{align}
The constraints give the limitation of the method, i.e., only under these
conditions we can proceed further. On the other hand, the constraints show the
relation between the spiral angle and the anisotropic exchange interaction. As
a result of Eq. (\ref{delta1}) and Eq. (\ref{delta2}), $H_{rot}$ becomes an
isotropic Hamiltonian
\begin{align}
H_{iso}= &  \sum_{j}J_{1}\Delta_{1}[\vec{s}_{j}\cdot\vec{s}_{j+1}%
+D_{1}^{\prime}\vec{s}_{j}\times\vec{s}_{j+1}\cdot\hat{z}]\nonumber\\
&  +J_{2}\Delta_{2}[\vec{s}_{j}\cdot\vec{s}_{j+2}+D_{2}^{\prime}\vec{s}%
_{j}\times\vec{s}_{j+2}\cdot\hat{z}]\label{h-rot-iso}%
\end{align}
where $D_{1}^{\prime}=\tan(\phi_{1}-\phi^{\prime}),D_{2}^{\prime}=\tan
(\phi_{2}-2\phi^{\prime})$.

As the anisotropy exchange interaction is \textquotedblleft rotated
\textquotedblright away for both NN and NNN interaction, it is readily shown
that $|FP,\hat{n}\rangle$ is an eigen state of $H_{iso}$ with the relation
\[
\sum_{i}\vec{s}_{i}\times\vec{s}_{j}\cdot\hat{z}|FP;\hat{n}\rangle=0.
\]
To prove the above equation, one only has to rotate $\hat{n}$ to $\hat{z}$ and
$\hat{z}$ to another direction. A more detailed analysis will be given in Sec.
\ref{symmetry} where the symmetry of the system is also discussed. As a
result,
\begin{equation}
H_{iso}|FP,\hat{n}\rangle=E_{0}|FP,\hat{n}\rangle\label{iso-state-n}%
\end{equation}
where $E_{0}=N(J_{1}\Delta_{1}+J_{2}\Delta_{2})/4$. In fact, Eqs. (4) and (5)
combined is the requirement that NN and NNN exchange interaction with
anisotropy can be transformed into isotropic exchange interactions simultaneously.


\section{Spiral state as the ground state\label{localsite}}

In this section, we will identify the region of FP state as the GS in
$H_{iso}$ (correspond to the spiral state as GS in $H$ ) by decomposing the
Hamiltonians into local Hamiltonians and applying positive semi-definite
theorem for analyzing. This method\cite{CSH} had been applied to zigzag spin
chains without DM interaction. It can also be applied to spin of any length.
Here we focus on spin-$1/2$ system.

\begin{figure}[h]
\begin{center}
\includegraphics[scale=0.55]{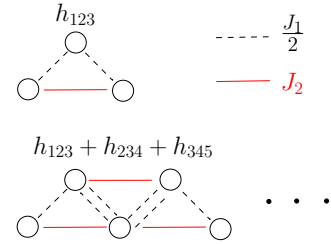}
\end{center}
\caption{The Hamiltonian can be decomposed into local Hamiltonians, and each
local Hamiltonian contains only the interactions between three neighboring
spins.}%
\label{ltog}%
\end{figure}

We dissect the Hamiltonian into many local Hamiltonians as shown in Fig.
\ref{ltog}. Each local Hamiltonian contains three spins. Thus, the original
Hamiltonian can be written as
\begin{align}
H_{iso} &  =\sum_{j=1}^{N}h_{j,j+1,j+2}\nonumber\\
&  =h_{1,2,3}\otimes\hat{1}_{2^{N-3}}+\sum_{j=2}^{N-2}\hat{1}_{2^{j-1}}\otimes
h_{j,j+1,j+2}\otimes\hat{1}_{2^{N-2-j}}\nonumber\\
&  +\hat{1}_{2^{N-3}}\otimes h_{N-1,N,1}+\hat{1}_{2^{N-3}}\otimes
h_{N,1,2}\label{expand-h}%
\end{align}
where the local Hamiltonian is giving by
\begin{align}
h_{j,j+1,j+2} &  =\frac{J_{1}\Delta_{1}}{2}[(\vec{s}_{j}+\vec{s}_{j+2}%
)\cdot\vec{s}_{j+1}]+J_{2}\Delta_{2}\vec{s}_{j}\cdot\vec{s}_{j+2}\nonumber\\
&  +\frac{J_{1}\Delta_{1}D_{1}^{\prime}}{2}[\vec{s}_{j}\times\vec{s}%
_{j+1}+\vec{s}_{j+1}\times\vec{s}_{j+2}]\cdot\hat{z}\nonumber\\
&  +J_{2}\Delta_{2}D_{2}^{\prime}\vec{s}_{j}\times\vec{s}_{j+2}\cdot\hat
{z}\label{localH}%
\end{align}
and $\hat{1}_{M}$ denotes the identity matrix of rank $M$. The direct product
of $h_{j,j+1,j+2}$ and $\hat{1}_{M}$ is meant to enlarge the vector space to
$2^{N}$ dimension to accommodate $N$ spins. The direct sum gives a Hamiltonian
matrix of $2^{N}$ dimension, as shown in Fig. \ref{ltog}. If FP state of three
spins is an eigen state of $h_{j,j+1,j+2}$, then that of $N$ spins is an eigen
state of $H_{iso}$. Furthermore, if a state of three spins has the lowest
energy under $H_{j,j+1,j+2}$, then corresponding state (constructed by direct
product) is the GS of $H_{iso}$. This is implied by the theorem of positive
semi-definite matrix discussed below.

To find the region for a FP state with spin direction $\widehat{n}$\ as the
ground state of Eq. (\ref{expand-h}), we rotate the $z$-axis to direction
$\hat{n}^{\prime}=(\sin(\beta)\cos(-\alpha),\sin(\beta)\sin(-\alpha
),\cos(\beta))$ and $\hat{n}$ to $\hat{z}$. The Hamiltonian of $H_{iso}$ in
Eq. (\ref{localH}) becomes
\begin{align}
h_{j,j+1,j+2}^{\hat{n}^{\prime}} &  =\frac{J_{1}\Delta_{1}}{2}[(\vec{s}%
_{j}+\vec{s}_{j+2})\cdot\vec{s}_{j+1}]+J_{2}\Delta_{2}\vec{s}_{j}\cdot\vec
{s}_{j+2}\nonumber\\
&  +\frac{J_{1}\Delta_{1}D_{1}^{\prime}}{2}[\vec{s}_{j}\times\vec{s}%
_{j+1}+\vec{s}_{j+1}\times\vec{s}_{j+2}]\cdot\hat{n^{\prime}}\nonumber\\
&  +J_{2}\Delta_{2}D_{2}^{\prime}\vec{s}_{j}\times\vec{s}_{j+2}\cdot
\hat{n^{\prime}}.\nonumber
\end{align}
and
\begin{align}
&  h_{j,j+1,j+2}^{\hat{n}^{\prime}}|\uparrow\rangle_{j}|\uparrow\rangle
_{j+1}|\uparrow\rangle_{j+2}\nonumber\\
&  =s^{2}(J_{1}\Delta_{1}+J_{2}\Delta_{2})|\uparrow\rangle_{j}|\uparrow
\rangle_{j+1}|\uparrow\rangle_{j+2}\nonumber\\
&  +\frac{\sqrt{2s}s}{4i}\sin(\beta)e^{-i\alpha}(\frac{J_{1}\Delta_{1}%
D_{1}^{\prime}}{2}+J_{2}\Delta_{2}D_{2}^{\prime})\nonumber\\
&  \times\left(  |\uparrow\rangle_{j}|\uparrow\rangle_{j+1}|\downarrow
\rangle_{j+2}-|\downarrow\rangle_{j}|\uparrow\rangle_{j+1}|\uparrow
\rangle_{j+2}\right)  \nonumber
\end{align}
where $D_{1}^{\prime}=\tan(\phi_{1}-\phi^{\prime}),D_{2}^{\prime}=\tan
(\phi_{2}-2\phi^{\prime})$. It is found that for $|\uparrow\rangle
_{j}|\uparrow\rangle_{j+1}|\uparrow\rangle_{j+2}$ to be an eigen state, the
required relation is
\begin{equation}
\frac{J_{1}\Delta_{1}}{J_{2}\Delta_{2}}=\frac{-2D_{2}^{\prime}}{D_{1}^{\prime
}}.\label{j1j2}%
\end{equation}
This procedure is not restricted to spin one-half systems. Substituting Eq.
(\ref{j1j2}) into Eq. (\ref{localH}), the local Hamiltonian becomes
\begin{align}
h_{j,j+1.j+2}=J_{2}\Delta_{2} &  [-\frac{D_{2}^{\prime}}{D_{1}^{\prime}}%
(\vec{s}_{j}+\vec{s}_{j+2})\cdot\vec{s}_{j+1}+\vec{s}_{j}\cdot\vec{s}%
_{j+2}\nonumber\\
&  -D_{2}^{\prime}(\vec{s}_{j}\times\vec{s}_{j+1}+\vec{s}_{j+1}\times\vec
{s}_{j+2})\cdot\hat{z}\nonumber\\
&  +D_{2}^{\prime}\vec{s}_{j}\times\vec{s}_{j+2}\cdot\hat{z}]\label{psd-h}%
\end{align}

To find the region for FP state to be the ground state of this local
Hamiltonian, we applied the positive semi-definite theorem. The theorem of
positive semi-definite (PSD) matrix is described by the following: \textit{The
necessary and sufficient condition for a real symmetric matrix} $A$ \textit{to
be positive semi-definite is} $x^{T}Ax\geq0$ \textit{for all real vectors}
$x$. \textit{If} $M$ \textit{and} $N$ \textit{are positive semi-definite, then
the sum} $M+N$, \textit{the direct sum} $M\oplus N,$ \textit{and direct
product} $M\otimes N$ \textit{are also positive semi-definite.} Hence, if we
are able to prove that $(h_{i,i+1,i+2}-E_{0})$ is a PSD matrix with $E_{0}$
being the energy of the FP state, then the $H_{iso}-E_{0}$ is also a PSD
matrix. For spin 1/2 system, the Hamiltonian in Eq. (\ref{psd-h}) is
$2^{3}\times2^{3}$ matrix. Its energy spectrum is
\begin{align}
E_{0} &  \mbox{ with four fold degeneracy};\nonumber\\
E_{0}+\delta E_{1} &  \mbox{ with two fold degeneracy}\nonumber\\
E_{0}+\delta E_{2} &  \mbox{ with two fold degeneracy}\label{local-energy}%
\end{align}
where $E_{0}$ is the energy of FP state, and
\begin{align*}
\frac{\delta E_{1}}{J_{2}\Delta_{2}} &  =\frac{2D_{1}^{\prime}D_{2}^{\prime
}-D_{1}^{\prime2}-\sqrt{\left[  3D_{1}^{\prime2}D_{2}^{\prime2}+(D_{1}%
^{\prime}+D_{2}^{\prime2})\right]  D_{1}^{\prime2}}}{2D_{1}^{\prime2}}\\
\frac{\delta E_{2}}{J_{2}\Delta_{2}} &  =\frac{2D_{1}^{\prime}D_{2}^{\prime
}-D_{1}^{\prime2}+\sqrt{\left[  3D_{1}^{\prime2}D_{2}^{\prime2}+(D_{1}%
^{\prime}+D_{2}^{\prime2})\right]  D_{1}^{\prime2}}}{2D_{1}^{\prime2}}%
\end{align*}
To make local Hamiltonian a PSD matrix, one requires $\delta E_{1}\geq0$ and
$\delta E_{2}\geq0$. Therefore the conditions for PSD are
\begin{equation}
\Big\{%
\begin{array}
[c]{ll}%
\quad(1-D_{1}^{\prime2})D_{2}^{\prime2}\geq2D_{1}^{\prime}D_{2}^{\prime
}\mbox{ and }2D_{1}^{\prime}D_{2}^{\prime}\geq D_{1}^{\prime2} &
\mbox{for }J_{2}\Delta_{2}\geq0\\
\mbox{or} & \\
\quad(1-D_{1}^{\prime2})D_{2}^{\prime2}\leq2D_{1}^{\prime}D_{2}^{\prime
}\mbox{ and }2D_{1}^{\prime}D_{2}^{\prime}\leq D_{1}^{\prime2} &
\mbox{for }J_{2}\Delta_{2}\leq0.
\end{array}
\label{cond-psd-1d}%
\end{equation}
This is the end of our derivation. Summarizing briefly, Eq. (4) and Eq. (5)
are the conditions of FP states being the eigen states of $H_{ios}$ in Eq.
(3). Hence they are also the conditions of the spiral spin states being the
eigen states of the physical Hamiltonian in Eq. (1). On the other hand, Eq.
(10) and inequality (13) are the conditions of the spiral spin states being
the ground states of the physical Hamiltonian in Eq. (1).

\begin{figure}[h]
\begin{center}
\includegraphics[scale=0.8]{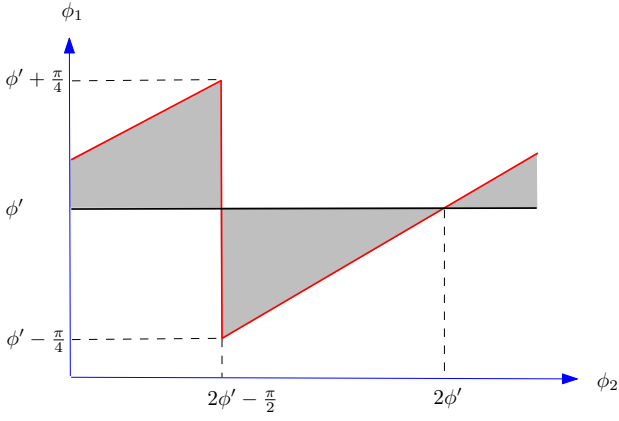}
\end{center}
\caption{The shaded regions show where spiral spin states are the ground
states and the relation between the pitch angle and $(\phi_{1},\phi_{2})$ in
the Hamiltonian, and the region for shifting $\pi$ in both $(\phi_{1}$ and
$\phi_{2})$ axes are also the PSD region.}%
\label{psd-region-p1-p2}%
\end{figure}

For most insulating compounds, $J_{2}$, the NNN superexchange interaction is
antiferromagnetic. Hence, we consider the case $J_{2}>0$ and $\Delta_{2}\geq
0$. The region for PSD in Eq. (\ref{cond-psd-1d}) can also be written as
\[
\Big\{%
\begin{array}
[c]{ll}%
2\phi^{\prime}-\frac{\pi}{2}\leq\phi_{2}\leq2\phi^{\prime} & \mbox{ for }\frac
{\phi_{2}}{2}\leq\phi_{1}\leq\phi^{\prime}\\
2\phi^{\prime}\leq\phi_{2}\leq2\phi^{\prime}+\frac{\pi}{2} & \mbox{ for }\phi
^{\prime}\leq\phi_{1}\leq\frac{\phi_{2}}{2}+\frac{\pi}{2}%
\end{array}
\]
It is indicated by the shaded area in Fig. \ref{psd-region-p1-p2}, for a given
$\phi^{\prime}$. Fig. \ref{psd-region-p1-p2} shows the main result of this
work. In view of Eq. (\ref{delta1}), Eq. (\ref{delta2}) and Eq. (\ref{j1j2}),
we see that there are two free parameters in Eq. (\ref{eq-H}), namely
$\phi_{1}(D_{1})$ and $\phi_{2}(D_{2})$. In the parameter space of $\phi_{1}$
and $\phi_{2}$, the shaded regions show where spiral spin states are the
ground states. The pitch angle $\phi^{\prime}$ of the spiral spins is closely
related to the values of $\phi_{1}(D_{1})$ and $\phi_{2}(D_{2})$. Hence, the
DM interaction plays the crucial role of determining the existence of spiral
spin states.

\begin{figure}[h]
\begin{center}
\includegraphics[scale=0.6]{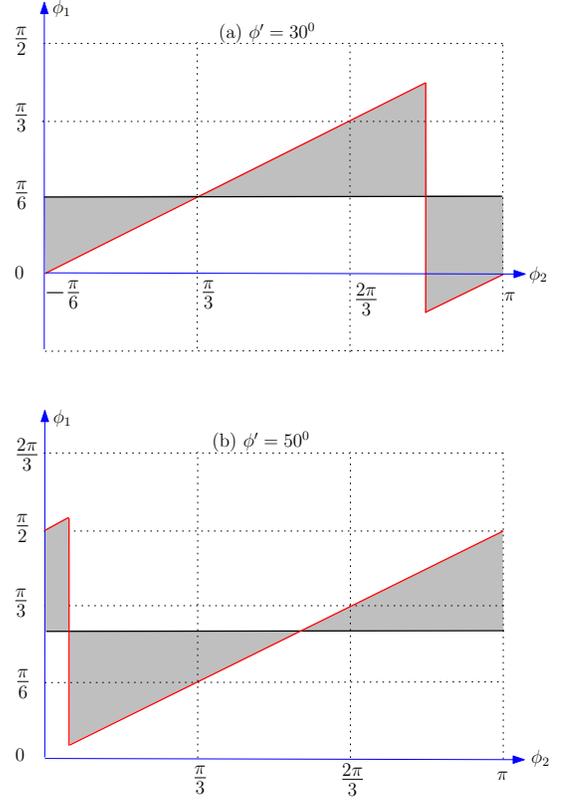}
\end{center}
\caption{The shaded regions show where spiral spin states are the ground
states and the relation between the pitch angle and $(\phi_{1},\phi_{2})$ in
the Hamiltonian, for (a)$\phi^{\prime}=30^{0}$, (b)$\phi^{\prime}=50^{0}$.}%
\label{example-psd-region}%
\end{figure}

It has been shown that inside the shaded region of Fig.
\ref{example-psd-region}, the spiral spin states are the ground state.
However, we also found that there are gapless excitations or Goldstone modes
(details are in sec. \ref{symmetry}). Outside the boundary, these modes
actually have lower energy as it will show in Eq. (\ref{symmetry-wk}). Hence,
the spiral spin states are the ground states in the shaded region. But they
are not the ground states outside of these boundaries which are exact since
our solution is exact. On the other hand, it is possible that the chiral
correlation or the in-plane spin correlation still exhibit long-range order
behavior outside of the boundaries. Furthermore, we cannot rule out the
possibility of the spiral spin states being the ground states in some other
region in phase space which is not in the neighborhood of the shaded regions.

For later use, we will show some cases explicitly in Fig.
\ref{example-psd-region}, the shaded regions will correspond to the condition
for spiral GS. In the following, we give two simple examples to illustrate our
result.

Similarly, for spin-s system of the Hamiltonian, the local site Hamiltonian is
a $(2s+1)^{3}\times(2s+1)^{3}$ matrix. After diagonalization, the energy
spectrum can be obtained. The theorem of PSD matrix can also give the region
for FP state as GS. Hence the result of this section can be generalized to
spins of any possible length.


\subsection{Isotropic exchange interaction\label{iso-h}}

For physical systems having isotropic superexchange, or $\Delta_{1}=\Delta
_{2}=1$ (in our case $\phi_{1}=\frac{\phi^{\prime}}{2},\phi_{2}=\phi^{\prime}%
$), the Hamiltonian of Eq.(\ref{eq-H}) becomes
\begin{align}
H^{\prime}=J_{2}  &  \sum_{i}-2\frac{\tan(\phi^{\prime})} {\tan(\frac
{\phi^{\prime}}{2})} \vec{s}_{i}\cdot\vec{s}_{i+1}+\vec{s}_{i}\cdot\vec
{s}_{i+2}\nonumber\\
&  -\tan(\phi^{\prime})(2\vec{s}_{i}\times\vec{s}_{i+1} -\vec{s}_{i}\times
\vec{s}_{i+2})\cdot\hat{z}\nonumber
\end{align}
The conditions for PSD of Eq.(\ref{cond-psd-1d}) turn out to be
\begin{equation}
\phi^{\prime}=\Big\{
\begin{array}
[c]{cc}%
0^{0}\sim90^{0},270^{0}\sim360^{0} & \mbox{ with }J_{2}>0,\\
90^{0}\sim270^{0} & \mbox{ with }J_{2}<0 .
\end{array}
\label{PSD-isotropic}%
\end{equation}
So in physical case $J_{2}>0$, the spiral angle $\phi^{\prime}$ is in the
range $0^{0}\sim90^{0}$ or $270^{0}\sim360^{0}$.

The implication of Eq. (\ref{PSD-isotropic}) can be seen by considering the
following three simple cases: (a) $\phi^{\prime}=0^{0}$, or the GS is the FP
state. In this case $H^{\prime}=J_{2}\sum_{i}-4 \vec{s}_{i}\cdot\vec{s}%
_{i+1}+\vec{s}_{i}\cdot\vec{s}_{i+2}$. It has been shown \cite{Nat88} that the
FP state and the UDRVB state are degenerate GS. Our method correctly leads to
one of them. (b) $\phi^{\prime}=\pi$, or the GS is the Neel state. In this
case $H^{\prime}=-|J_{2}|\sum_{i}\vec{s}_{i}\cdot\vec{s}_{i+2}$. The system
becomes two decoupled spin chains. Each is ferromagnetically coupled. (c) The
GS is spiral state with spiral angle being $30^{0}$. The resulting parameters
are $(J_{1},J_{2},\Delta_{1},\Delta_{2},D_{1},D_{2})=(-4.13,1,1,10.268,0.577)$%
. (d) For specific spiral angle $\phi^{\prime}=50^{0}$, the resulting
parameters are $(J_{1},J_{2},\Delta_{1},\Delta_{2},D_{1},D_{2}%
)=(-5.111,1,1,1,0.466,1.191)$.

In this isotropic Hamiltonian, the relative strength of exchange coupling is
$J_{1}/J_{2}=-2\tan(\phi^{\prime})/\tan(\phi^{\prime}/2)$. In the region of
PSD in Eq. (\ref{PSD-isotropic}) with $J_{2}>0$, this is always less than
$-4$. We shall see in Sec. \ref{extension} that for coupled spin chains or
two-dimensional cases, the superexchange $J_{1}/J_{2}$ can be tuned by inter
chain coupling $J_{3}$ to make $|J_{1}/J_{2}|$ smaller.


\subsection{Without DM interaction\label{nodm}}

For system containing high symmetry or small spin-orbital interaction, we have
the case $D_{1}\approx D_{2}\approx0$, implying $\phi_{1}=\phi_{2}=0$. Then
the Hamiltonian in Eq. (\ref{eq-H}) under the condition of Eqs. (\ref{delta1}%
,\ref{delta2},\ref{j1j2}) becomes
\begin{align}
H^{\prime\prime}=J_{2} &  \sum_{j}-4\cos(\phi^{\prime})(\cos(\phi^{\prime
})s_{j}^{z}s_{j+1}^{z}+s_{j}^{x}s_{j+1}^{x}+s_{j}^{y}s_{j+1}^{y})\nonumber\\
&  +(\cos(2\phi^{\prime})s_{j}^{z}s_{j+2}^{z}+s_{j}^{x}s_{j+2}^{x}+s_{j}%
^{y}s_{j+2}^{y})\label{nodm-eq}%
\end{align}
and the spiral state is the eigen state. Now we have
\begin{equation}
\frac{J_{1}}{J_{2}}=-4\cos(\phi^{\prime}),\label{nodm-eq-f}%
\end{equation}
which is consist with the result of treating spins as classic vectors.

Here are several examples of the system in Eq. (\ref{nodm-eq}): (a) For GS
with pitch angle $\phi^{\prime}=30^{0}$, the Hamiltonian has parameters
$(J_{1},J_{2},\Delta_{1},\Delta_{2},D_{1},D_{2})=(-3.464,1,0.866,0.5,0,0)$.
(b) For GS with specific pitch angle $\phi^{\prime}=50^{0}$, the Hamiltonian
has parameter $(J_{1},J_{2},\Delta_{1},\Delta_{2},D_{1},D_{2}%
)=(-2.571,1,0.643,-0.174,0,0)$.

The spiral spin state is always the ground state provided the anisotropy
interactions between NN and NNN have the forms in Eq. (\ref{nodm-eq}) with
$J_{2}>0$. However, this kind of solution may not be possible for
multiferroics LiCu$_{2}$O$_{2}$ with $\phi^{\prime}=62^{0}$, because in this
case $\Delta_{2}=\cos(124^{0})<0$, and it is likely to be unpractical. It will
be shown later, even if inter-chain coupling is considered, $\Delta_{2}$
remains negative. This suggest that there is DM interaction in LiCu$_{2}%
$O$_{2}$.


\subsection{Special case with $\phi^{\prime}=50^{0}$\label{case-50}}

For multiferroic compound LiCu$_{2}$O$_{2}$, the spiral angle is close to
$\pi/3$, therefore here we consider this special case. DM interaction and
anisotropy are likely to be small. The parameter set $\{(J_{1},J_{2}%
),(\Delta_{1},\Delta_{2}),(D_{1},D_{2})\}=\{(-3.111,1),(1,0),(0.176,0.176)\}$
gives us $\phi^{\prime}=50^{0}$, which is close to $\pi/3$. However, we note
that $\Delta_{2}$ vanishes in this case. We suggest that one can treat
$J_{2}\Delta_{2}s_{i}^{z}s_{i+1}^{z}$ as a perturbation. Since $J_{2}%
\Delta_{2}>0$, the perturbed state should have greater spiral angle. Hence,
our approach should to be capable of dealing with realistic physical system
such as LiCu$_{2}$O$_{2}$. The same approach can be used for other value of
spiral angle $\phi^{\prime}$.


\section{Symmetry properties\label{symmetry}}

In Sec. \ref{2connect}, we showed that FP state in Eq. (\ref{iso-state-n}) for
any direction $\hat{n}$ is an eigen state. Later in Sec. \ref{localsite}, we
identified the region for these states to be GS. These states form a Hilbert
space which has SU(2) symmetry. We can implement a site-dependent unitary
rotation (O(2) rotation) to generate spiral states. On the other hand, the
Hamiltonians $H$, $H_{rot}$ and $H_{iso}$ all have only $SO(2)$ symmetry.
Hence, we arrived at a situation which is called "emergent symmetry" by
Batista \cite{Ba09}. The symmetry group of the Hamiltonian in Eq. (\ref{eq-H})
is isomorphic to that of $H_{iso}$ which is $SO(2)$. However, the Hilbert
space of degenerate spiral states has a symmetry group which is isomorphic to
$SU(2)$. We give a detailed analysis in this section.

To see the symmetry property of this system more clearly, we first express our
results following the notations of Batista \cite{Ba09}. We note in passing
that our result contain DM interaction, which is a generalization. The
Hamiltonian of Eq.(\ref{eq-H}) in $k$-space is
\begin{align}
H=\sum_{q} &  [J_{1}\Delta_{1}\cos(q)+J_{2}\Delta_{2}\cos(2q)]s_{q}^{z}%
s_{-q}^{z}\nonumber\\
&  +[J_{1}(\cos(q)-D_{1}\sin(q))\nonumber\\
&  +J_{2}(\cos(2q)-D_{2}\sin(2q))]s_{q}^{+}s_{q}^{-}\label{h-k}%
\end{align}
where $s_{q}^{z}=\frac{1}{\sqrt{L}}\sum_{j}e^{iqj}s_{j}^{z},s_{q}^{+}=\frac
{1}{\sqrt{L}}\sum_{j}e^{iqj}s_{j}^{+}$, and $s_{q}^{-}=(s_{q}^{+})^{\dagger}$,
with lattice constant set to $1$. The commutation relation of $s_{q=\phi
^{\prime}}^{\dagger}$ is
\[
\lbrack s_{\phi^{\prime}}^{+},H]=\frac{i}{\sqrt{L}}\sum_{l}e^{il\phi^{\prime}%
}s_{l}^{+}a_{l}%
\]
where
\begin{align*}
a_{l} &  =\frac{J_{1}\sin(\phi_{1}-\phi^{\prime})}{\cos(\phi_{1})}(s_{l+1}%
^{z}-s_{l-1}^{z})\\
&  +\frac{J_{2}\sin(\phi_{2}-2\phi^{\prime})}{\cos(\phi_{2})}(s_{l+2}%
^{z}-s_{l-2}^{z})
\end{align*}
From $H|FP,-\hat{z}\rangle=E_{0}|FP,-\hat{z}\rangle$ and $a_{l}|FP,-\hat
{z}\rangle=0$, we get the relation of $H$ in Eq.(\ref{h-k}) as
\[
\lbrack s_{\phi^{\prime}}^{+},(H-E_{0})](s_{\phi^{\prime}}^{+})^{p}%
|FP,-\hat{z}\rangle=0
\]
for any integer $p$. Hence, $(s_{\phi^{\prime}}^{+})^{p}|FP,-\hat{z}\rangle$
is one of the degenerate eigen states of $H$. The spiral spin states is a
linear combination of $(s_{\phi^{\prime}}^{+})^{p}|FP,\hat{z}\rangle$, we show
it more directly below.

It is clearer to discuss the symmetry property by analyze the transformed
Hamiltonian. The FP states are the degenerate GS of $H_{iso}$ irrespective of
the direction of $\hat{n}$. To put it in another way, $H_{iso}$ commute with
$\sum_{i}s_{i}^{z}$ so that the $z$-component of the total spin is a good
quantum number, hence, they have the following eigen states
\begin{equation}
\{|0\rangle,\sum_{j}|j\rangle,\sum_{j<k}|j,k\rangle,\sum_{j<k<l}%
|j,k,l\rangle,\dots,|FP;\hat{z}\rangle\}\label{eq-FP}%
\end{equation}
where $|j,k,\dots\rangle\equiv S_{j}^{+}S_{k}^{+}\dots|0\rangle$ and
$|0\rangle\equiv|FP;-\hat{z}\rangle$. Since any FP state $|FP,\hat{n}\rangle$
is a linear combination of the above states, the states in Eq. (\ref{eq-FP})
are actually degenerate GS of $H_{iso}$. Under the unitary transformation, the
above states can be transformed into magnon states:
\[
\{|0\rangle,s_{\phi^{\prime}}^{+}|0\rangle,(s_{\phi^{\prime}}^{+}%
)^{2}|0\rangle,(s_{\phi^{\prime}}^{+})^{3}|0\rangle,\dots,|FP;\hat{z}\rangle\}
\]
The states $(s_{\phi^{\prime}}^{+})^{p}|FP,-\hat{z}\rangle$, those in
Eq.(\ref{eq-FP}) and $|FP;\hat{n}\rangle$ are three basis of the degenerate GS
of the Hamiltonian in Eq. (\ref{eq-H}). Each set form a Hilbert space with a
symmetry group isomorphic to SU(2) which contains the symmetry group
(isomorphic to SO(2)) of the Hamiltonian.

As for the excitation energy of the Hamiltonian in Eq. (\ref{eq-H}), consider
one magnon with wave vector $k$ as $|\psi\rangle=\sum_{j}e^{ikj}|j\rangle$.
Using the relation
\begin{align*}
&  (H-NE_{0})|j\rangle=-(J_{1}\Delta_{1}+J_{2}\Delta_{2})|j\rangle\\
&  +\frac{J_{1}}{2}[(1+iD_{1})|j-1\rangle+(1-iD_{1})|j+1\rangle]\\
&  +\frac{J_{2}}{2}[(1+iD_{2})|j-2\rangle+(1-iD_{2})|j+2\rangle]
\end{align*}
we found that $|\psi\rangle$ is an eigen state and its energy spectrum is
\begin{align}
w_{k} &  =E-NE_{0}\nonumber\\
&  =J_{1}(\cos(k)-\Delta_{1}+D_{1}\sin(k))\nonumber\\
&  +J_{2}(\cos(2k)-\Delta_{2}+D_{2}\sin(2k)).\label{symmetry-wk}%
\end{align}
When $k$ approaches $\phi^{\prime}$, the energy $E$ converges to $E_{0}$.
Hence we have gapless excitation. Note that the one-magnon states are not the
only low-lying excitations. In fact, there are certain two-magnon,
multiple-magnon states that are also gapless excitation. Studying on this
topic will be presented in another paper.


\section{Extension of the model\label{extension}}

Our result can be generalized to the cases of coupled spin chains and certain
types of two-dimensional triangular lattices. Hence, physical systems such as
LiCu$_{2}$O$_{2}$ and ACrO$_{2}$(A=Cu,Ag,Li or Na) can be studied with our method.


\subsection{Coupled spin-1/2 zigzag spin system\label{2d-zig}}

\begin{figure}[h]
\begin{center}
\includegraphics[scale=0.7]{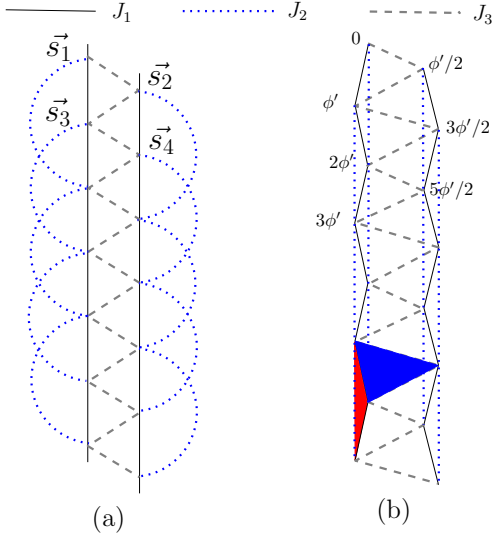}
\end{center}
\caption{(a) Two coupled zigzag spin chains. (b) Two basic elements denoted as
a blue (light) triangular and a red (dark) triangular for two coupled spin
chains.}%
\label{two-leg-zigzag}%
\end{figure}
Recently, the compound LiCu$_{2}$O$_{2}$ \cite{li08,li09} attracted
researchers' attention because of its multiferroic property. It has chains of
edge-sharing oxygen plaquettes with copper ions at the centers. The NN and NNN
superexchange interactions strength between copper ions are comparable. A
coupled zigzag spin ladder with two legs are shown in Fig.
\ref{two-leg-zigzag}(a). The model Hamiltonian is given by
\begin{align}
H^{csc}= &  \sum_{j}J_{1}[\Delta_{1}s_{j}^{z}s_{j+2}^{z}+\frac{1}{2}(s_{j}%
^{+}s_{j+2}^{-}+s_{j}^{-}s_{j+2}^{+})\nonumber\\
&  +D_{1}\vec{s}_{j}\times\vec{s}_{j+2}\cdot\hat{z}]\nonumber\\
&  +J_{2}[\Delta_{2}s_{j}^{z}s_{j+4}^{z}+\frac{1}{2}(s_{j}^{+}s_{j+4}%
^{-}+s_{j}^{-}s_{j+4}^{+})\nonumber\\
&  +D_{2}\vec{s}_{j}\times\vec{s}_{j+4}\cdot\hat{z}]\nonumber\\
&  +J_{3}[\Delta_{3}s_{j}^{z}s_{j+1}^{z}+\frac{1}{2}(s_{j}^{+}s_{j+1}%
^{-}+s_{j}^{-}s_{j+1}^{+})\nonumber\\
&  +D_{3}\vec{s}_{j}\times\vec{s}_{j+1}\cdot\hat{z}]\label{2d-hp}%
\end{align}
where $j$ is the label of the lattice site, $J_{1}$ is the nearest neighbor
(NN) interaction, $J_{2}$ the next nearest neighbor (NNN) interaction, $J_{3}$
is the inter-chain coupling, $\Delta_{1}(\Delta_{2},\Delta_{3})$ is the
anisotropic interaction along $z$ axis for NN (NNN, inter-chain) interaction
and $D_{i}^{\prime}s$ are the respective strength of DM interaction.
Superscript $csc$ denotes coupled spin chain. With similar procedure in
Sec.\ref{2connect}, we set $D_{1}=\tan(\phi_{1}),D_{2}=\tan(\phi_{2}%
),D_{3}=\tan(\phi_{3})$, and construct the unitary transformation rotating the
spins around $z$-axis: $U=\prod_{j=1}^{N}\exp(is_{j}^{z}\vec{Q}\cdot\vec
{R}_{j})$ with a constant $\phi^{\prime}=2\vec{Q}\cdot(\vec{R}_{j+1}-\vec
{R}_{j})$. The spiral angle is shown in Fig. \ref{two-leg-zigzag}(b). This
gives a fixed phase difference between two chains. The conditions of finding a
unitary transformation $U(\phi^{\prime})$ to change $H$ into an isotropic
Hamiltonian are
\begin{align*}
\Delta_{1} &  =\frac{\cos(\phi_{1}-\phi^{\prime})}{\cos(\phi_{1})},\\
\Delta_{2} &  =\frac{\cos(\phi_{2}-2\phi^{\prime})}{\cos(\phi_{2})},\\
\Delta_{3} &  =\frac{\cos(\phi_{3}-\frac{\phi^{\prime}}{2})}{\cos(\phi_{3})}.
\end{align*}
The resulting isotropic Hamiltonian is
\begin{align*}
H_{iso}^{csc}= &  \sum_{j}J_{1}\Delta_{1}[\vec{s}_{j}\cdot\vec{s}_{j+2}%
+D_{1}^{\prime}\vec{s}_{j}\times\vec{s}_{j+2}\cdot\hat{z}]\\
&  +J_{2}\Delta_{2}[\vec{s}_{j}\cdot\vec{s}_{j+4}+D_{2}^{\prime}\vec{s}%
_{j}\times\vec{s}_{j+4}\cdot\hat{z}]\\
&  +J_{3}\Delta_{3}[\vec{s}_{j}\cdot\vec{s}_{j+1}+D_{3}^{\prime}\vec{s}%
_{j}\times\vec{s}_{j+1}\cdot\hat{z}]
\end{align*}
where $D_{1}^{\prime}=\tan(\phi_{1}-\phi^{\prime}),D_{2}^{\prime}=\tan
(\phi_{2}-2\phi^{\prime})$, and $D_{3}^{\prime}=\tan(\phi_{3}-\frac
{\phi^{\prime}}{2})$. If for Hamiltonian $H_{iso}^{csc}$, FP states are the
eigen states then for Hamiltonian $H^{csc}$, spiral states are the eigen states.

\begin{figure}[h]
\begin{center}
\includegraphics[scale=0.6]{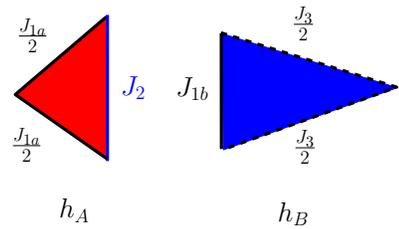}
\end{center}
\caption{The two basic elements of the system:red (dark) triangles $h_{A}$ and
blue (light) triangles $h_{B}$. }%
\label{two-local-site}%
\end{figure}
We again dissect $H_{iso}^{csc}$ into local Hamiltonians, each contains three
spins. However, now there are two kinds of local Hamiltonians, represented by
blue (light) and red (dark) triangles in Fig. \ref{two-leg-zigzag}(b). Since
the two triangles have a common edge, we divide $J_{1}$ into $J_{1a}$ and
$J_{1b}$. This division is arbitrary. As shown in Fig. \ref{two-local-site},
the two local Hamiltonians are $h_{A}$ (dark or red)and $h_{B}$ (light or
blue:)
\begin{align}
h_{A} &  =\frac{J_{1a}\Delta_{1}}{2}[(\vec{s}_{i}+\vec{s}_{i+4})\cdot\vec
{s}_{i+2}]+J_{2}\Delta_{2}\vec{s}_{i}\cdot\vec{s}_{i+4}\nonumber\\
&  +\frac{J_{1a}\Delta_{1}D_{1}^{\prime}}{2}[\vec{s}_{i}\times\vec{s}%
_{i+2}+\vec{s}_{i+2}\times\vec{s}_{i+4}]\cdot\hat{z}\nonumber\\
&  +J_{2}\Delta_{2}D_{2}^{\prime}\vec{s}_{i}\times\vec{s}_{i+4}\cdot\hat
{z},\label{2d-ha}\\
h_{B} &  =\frac{J_{3}\Delta_{3}}{2}[(\vec{s}_{i}+\vec{s}_{i+2})\cdot\vec
{s}_{i+1}]+J_{1b}\Delta_{1}\vec{s}_{i}\cdot\vec{s}_{i+2}\nonumber\\
&  +\frac{J_{3}\Delta_{3}D_{3}^{\prime}}{2}[\vec{s}_{i}\times\vec{s}%
_{i+1}+\vec{s}_{i+1}\times\vec{s}_{i+2}]\cdot\hat{z}\nonumber\\
&  +J_{1b}\Delta_{1}D_{1}^{\prime}\vec{s}_{i}\times\vec{s}_{i+2}\cdot\hat
{z}.\label{2d-hb}%
\end{align}
$|FP,\hat{n}\rangle$ will be the eigen state locally, provided the following
conditions which are analogous to that in Eq. (\ref{j1j2}), are satisfied by
\begin{align}
\frac{J_{1a}}{J_{2}} &  =\frac{-2D_{2}^{\prime}\Delta_{2}}{D_{1}^{\prime
}\Delta_{1}},\label{2d-c-1}\\
\frac{J_{3}}{J_{1b}} &  =\frac{-2D_{1}^{\prime}\Delta_{1}}{D_{3}^{\prime
}\Delta_{3}}.\label{2d-c-2}%
\end{align}
The forms of both local Hamiltonians $h_{A}$ and $h_{B}$ in Eq. (\ref{2d-ha})
and Eq. (\ref{2d-hb}) are the same as those of Eq. (\ref{psd-h}). Hence the
spectrum in Eq. (\ref{local-energy}) is applicable. The energy difference
between the FP state and others for $h_{A}$ and $h_{B}$ are
\begin{align*}
\frac{\delta E_{A}^{\pm}}{J_{2}\Delta_{2}} &  =\frac{2D_{1}^{\prime}%
D_{2}^{\prime}-D_{1}^{\prime2}\pm\sqrt{\lbrack3D_{2}^{\prime2}D_{1}^{\prime
2}+(D_{1}^{\prime}+D_{2}^{\prime})^{2}]D_{1}^{\prime2}}}{2D_{1}^{\prime2}}\\
\frac{\delta E_{B}^{\pm}}{J_{1b}\Delta_{1}} &  =\frac{2D_{1}^{\prime}%
D_{3}^{\prime}-D_{3}^{\prime2}\pm\sqrt{\lbrack3D_{1}^{\prime2}D_{2}^{\prime
2}+(D_{3}^{\prime}+D_{1}^{\prime})^{2}]D_{3}^{\prime2}}}{2D_{3}^{\prime2}}%
\end{align*}
For $H-E_{0}$ to be a PSD matrix, one requires that $\delta E_{A}^{\pm}\geq0$
and $\delta E_{B}^{\pm}\geq0$. Therefore the conditions for PSD are
\begin{widetext}
\begin{eqnarray*}
&&\qquad\Big\{\begin{array}{ll}
\quad (1-D_1^{\prime 2})D_2^{\prime 2}\geq 2D_1^{\prime }D_2^{\prime }\mbox{ and }
2D_2^{\prime }\geq D_1^{\prime } & \mbox{ for }J_2\Delta_2\geq0, \\
\mbox{or}\\
\quad(1-D_1^{\prime 2})D_2^{\prime 2}\leq 2D_1^{\prime }D_2^{\prime }\mbox{ and }
2D_2^{\prime }\leq D_1^{\prime } & \mbox{ for }J_2\Delta_2\leq0
\end{array} \\
&&\mbox{and}\\
&&\qquad\Big\{\begin{array}{ll}
\quad(1-D_3^{\prime 2})D_1^{\prime 2}\geq 2D_3^{\prime }D_1^{\prime }\mbox{ and }
2D_1^{\prime }\geq D_3^{\prime } & \mbox{ for }J_{1b}\Delta_1\geq0, \\
\mbox{or}\\
\quad(1-D_3^{\prime 2})D_1^{\prime 2}\leq 2D_3^{\prime }D_1^{\prime }\mbox{ and }
2D_1^{\prime }\leq D_3^{\prime } & \mbox{ for }J_{1b}\Delta_1\leq0. \\
\end{array}
\end{eqnarray*}
\end{widetext}

The implication of above derivation can be seen by the cases similar to those
of Sec.\ref{iso-h}, Sec.\ref{nodm} and Sec.\ref{case-50}

\begin{itemize}
\item For the isotropic case with $\Delta_{1}=\Delta_{2}=\Delta_{3}=1$, and
hence $\phi_{1}=\phi^{\prime}/2,\phi_{2}=\phi^{\prime},\phi_{3}=\phi^{\prime
}/4$, the Hamiltonian is
\begin{align*}
H^{csc}=\sum_{i}  &  [\frac{-2}{\tan(\frac{\phi^{\prime}}{2})}(J_{2}\tan
(\phi^{\prime})+J_{3}\tan(\frac{\phi^{\prime}}{4})]\vec{s}_{i}\cdot\vec
{s}_{i+2}\\
&  +J_{2}\vec{s}_{i}\cdot\vec{s}_{i+4}+J_{3}\vec{s}_{i}\cdot\vec{s}_{i+1}\\
&  +2[J_{2}\tan(\phi^{\prime})+J_{3}\tan(\frac{\phi^{\prime}}{4})]\vec{s}%
_{i}\times\vec{s}_{i+2}\cdot\hat{z}\\
&  +J_{2}\tan(\phi^{\prime})\vec{s}_{i}\times\vec{s}_{i+4}\cdot\hat{z}\\
&  +J_{3}\tan(\frac{\phi^{\prime}}{4})\vec{s}_{i}\times\vec{s}_{i+1}\cdot
\hat{z}%
\end{align*}
The resulting relations between $J_{1}$ and $J_{2},J_{3}$ from Eq.
(\ref{2d-c-1}) and Eq. (\ref{2d-c-2}) are
\begin{equation}
J_{1}=\frac{-2}{\tan(\frac{\phi^{\prime}}{2})}[J_{2}\tan(\phi^{\prime}%
)+J_{3}\tan(\frac{\phi^{\prime}}{4})].
\end{equation}
Similar to Eq. (\ref{PSD-isotropic}), we get the PSD conditions for $h_{A}$ in
Eq. (\ref{2d-ha}) and $h_{B}$ in Eq. (\ref{2d-hb})
\begin{align*}
\phi^{\prime}  &  =\Big\{%
\begin{array}
[c]{cc}%
0^{0}\sim90^{0},270^{0}\sim360^{0} & \mbox{with }J_{2}>0\\
90^{0}\sim270^{0} & \mbox{with }J_{2}<0
\end{array}
\\
&  \mbox{and }\quad\\
\phi^{\prime}  &  =\Big\{%
\begin{array}
[c]{cc}%
0^{0}\sim180^{0},540^{0}\sim720^{0} & \mbox{with }J_{1b}>0\\
180^{0}\sim540^{0} & \mbox{with }J_{1b}<0
\end{array}
\end{align*}
Note that $\vec{Q}\cdot(\vec{R}_{i+1}-\vec{R}_{i})=\phi^{\prime}/2$, so
$\phi^{\prime}$ and $\phi^{\prime}+2\pi$ do not give the same spiral spin
state. For antiferromagnetic coupling ($J_{2}>0$), and no constraint on
$J_{1b}$, the PSD condition requires the spiral angle $\phi^{\prime}$ to be in
the region $0^{0}\sim90^{0}$ or $270^{0}\sim360^{0}$. We give the following
examples. For spiral angle $\phi^{\prime}=30^{0}$, and $J_{3}/J_{2}=\pm0.3$,
the following set of parameters satisfy PSD condition: $\{(J_{1},J_{2}%
,J_{3}),(D_{1},D_{2},D_{3})\}=\{(-3.772,1,-0.3),(0.268,0.577,0.132)\}$ with
positive $J_{2}$ and $J_{1b}$.

\item For the case of no DM interaction as that in Sec. \ref{nodm}, we again
give the example of $\phi^{\prime}=30^{0}, J_{3}/J_{2}=\pm0.3$. The parameters
satisfying PSD conditions can be $\{(J_{1},J_{2},J_{3}),(\Delta_{1},\Delta
_{2},\Delta_{3})\}=\{(3.386,1,-0.3),(0.866,0.5,0.966)\}$ with positive $J_{2}$
and $J_{1b}$.

\item For the case $\phi^{\prime}=50^{0}$, we set $\phi_{1}=\phi_{2}=\phi
_{3}=10^{0}$ and $J_{3}/J_{2}=\pm0.3$, and get the parameter set of the system
$\{(J_{1},J_{2},J_{3}),(\Delta_{1},\Delta_{2},\Delta_{3}),(D_{1},D_{2}%
,D_{3})\}$ as $\{(-3.151,1,-0.3),(0.778,0,1),(0.176,0.176,0.176)\}$ with
positive $J_{2}$ and $J_{1b}$. Above parameters are very close to the
realistic systems, For example, in LiCu$_{2}$O$_{2}$, the exchange
interactions given by ab initio calculation \cite{Gipp04} are very close to
our parameters. The DM interaction we get are smaller than $J_{2}$ by an order
of magnitude. The only possible discrepancy is $\Delta_{2}$ which should be
close to $1$. We suggest the term $J_{2}\Delta_{2} s_{i}^{z}s_{i+2}^{z}$
should be treated as a perturbation.
\end{itemize}


\subsection{2D triangular spin-$1/2$ spin system with spiral state
\label{2d-triangular}}

\begin{figure}[h]
\begin{center}
\includegraphics[scale=0.75]{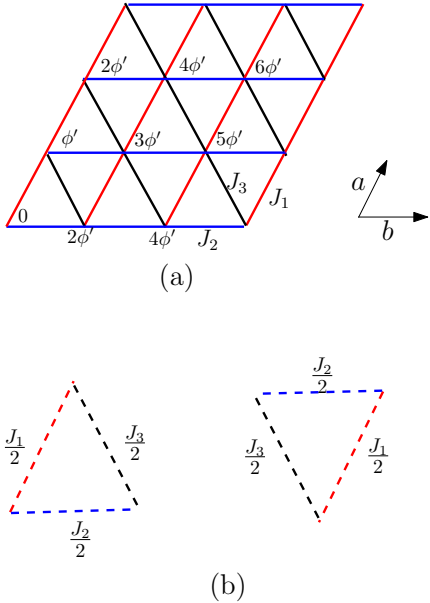}
\end{center}
\caption{(a) Two-dimensional triangular lattice system containing $J_{1}%
$(red), $J_{3}$(black) and $J_{2}$(blue) superexchange interaction, with the
spiral angle at different sites are shown. (b) Two basic elements of the
triangular lattice.}%
\label{triangular-lattice}%
\end{figure}
Our method can be applied to the compounds of layers of triangular lattice,
such as ACrO$_{2}$ \cite{To08}(A=Cu, Ag, Li or Na) shown in Fig.
\ref{triangular-lattice}(a). Here we assume that the exchange interaction
$J_{1}$ along $a$-axis,$J_{2}$ along $b$-axis is $J_{2}$ and $J_{3}$ along
$\vec{a}-\vec{b}$ axis. For simplicity, we consider the case $J_{3}=J_{1}$,
then we have
\begin{align}
H^{2D}=  &  \sum_{\langle j,k\rangle}J_{1}[\Delta_{1}s_{j,k}^{z}s_{j^{\prime
},k}^{z} +\frac{1}{2}(s_{j,k}^{+}s_{j^{\prime},k}^{-}+s_{j,k}^{-}s_{j^{\prime
},k}^{+})\nonumber\\
&  +D_{1}\vec{s}_{j,k}\times\vec{s}_{j^{\prime},k}\cdot\hat{z}]\nonumber\\
&  +J_{2}[\Delta_{2}s_{j,k}^{z}s_{j,k^{\prime}}^{z}+\frac{1}{2} (s_{j,k}%
^{+}s_{j,k^{\prime}}^{-}+s_{j,k}^{-}s_{j,k^{\prime}}^{+})\nonumber\\
&  +D_{2}\vec{s}_{j,k}\times\vec{s}_{j,k^{\prime}}\cdot\hat{z}]]
\label{2d-trian-h}%
\end{align}
where $j$ stands for the site index along $b$-direction, and $k$ is the index
along $a$-direction. Defining $D_{1}=\tan(\phi_{1}),D_{2}=\tan(\phi_{2})$, and
performing a unitary transformation $U(\phi^{\prime})$, we get an isotropic
Hamiltonian
\begin{align*}
H_{iso}^{2D}=  &  \sum_{\langle j,k\rangle}J_{1}\Delta_{1} [\vec{s}_{j,k}%
\cdot\vec{s}_{j^{\prime},k}+D_{1}^{\prime}\vec{s}_{j,k} \times\vec
{s}_{j^{\prime},k}\cdot\hat{z}]\\
&  +J_{2}\Delta_{2} [\vec{s}_{j,k}\cdot\vec{s}_{j,k^{\prime}} +D_{2}^{\prime
}\vec{s}_{j,k}\times\vec{s}_{j,k^{\prime}}\cdot\hat{z}]
\end{align*}
with the parameters $\Delta_{1}=\frac{\cos(\phi_{1}-\phi^{\prime})}{\cos
(\phi_{1})}, \Delta_{2}=\frac{\cos(\phi_{2}-2\phi^{\prime})}{\cos(\phi_{2})}$
and $D_{1}^{\prime}=\tan(\phi_{1}-\phi^{\prime}),D_{2}^{\prime} =\tan(\phi
_{2}-2\phi^{\prime})$. For Hamiltonian $H_{iso}^{2D}$ , FP state is the eigen
state, and for Hamiltonian $H^{2D}$, spiral state is the eigen state.

The Hamiltonian of a layer is the combination of many local Hamiltonians of
small triangles, as shown in Fig. \ref{triangular-lattice}(b). For each local
Hamiltonian, we can perform our analysis as before. The local site Hamiltonian
is the same as that described in 1D zigzag spin chain. Therefore, the region
for PSD is the same as that of zigzag spin chain in Eq. (\ref{cond-psd-1d}),
and Eq. (\ref{j1j2}) also applies.

Consider isotropic exchange interaction as an example. We set $\Delta
_{1}=\Delta_{2}=1$ and $\phi_{1}=\phi^{\prime}/2,\phi_{2}=\phi^{\prime}$. It
was found that the PSD condition gives
\[
\phi^{\prime}=\Big\{
\begin{array}
[c]{cc}%
0^{0}\sim90^{0},270^{0}\sim360^{0} & ,\mbox{ for }J_{2}>0\\
90^{0}\sim270^{0} & ,\mbox{ for }J_{2}<0
\end{array}
\]
Our method has the potential application to two-dimensional spin system.


\section{Analysis and discussion\label{compare}}

We now compare our results with those of numerical calculations and
simulations. The results from the calculations (see below) show that the
zigzag spin chain possesses rather rich and sophisticate physics. Since our
method is to find the exact solutions, we can only access limited regions in
phase space. However, inside these regions, our results are exact and thus can
be used as standard to check the results of numerical calculations. Our
calculation can also serve as a guideline for the calculation of the
correlation functions outside of these regions. Hence, by comparison, we hope
to shed some light on the complex physical landscape of the zigzag spin chain.

There are many works on the phase diagram. Hikihara et al. \cite{Hik01} used
Density matrix renormalization group (DMRG) method and found spin-liquid,
dimer and gapless chiral phase for various quantum spins. For $s=1/2$, they
found notably, the chiral correlation function $\langle\vec{s}_{j}\times
\vec{s}_{j+d}\rangle$, has long-range order behavior. Plekhanov et
al.\cite{Ple10} also used DMRG technique to study similar systems. They found
that quantum fluctuation modified the classical phase diagram. In particular,
they found phases denoted by spin liquid I, spin liquid II, E-I and E-II. In
the former two phases, they found that the spin correlation functions
(especially those on x-y plane $\langle s_{j}^{x}s_{j+d}^{x}\rangle$) have the
power-law decay behavior. In the E-I and E-II phases, the spin correlation
functions decrease exponentially except for a weak magnetization in E-I phase.
Jafari et al.\cite{Jaf07} used quantum renormalization group, to find a
similar phase diagram except for a new phase near the spin liquid I. It was
designated as dimer II. For $J_{1} <0$, the spin liquid III have more or less
occupied regions of E-II phase of Plekhanov et al.\cite{Ple10}. Dmitriev and
Krivnov \cite{Dv07} studied the same problem with variational mean-field
approximation and scaling. They also obtained similar region of spin-liquid I
in the phase diagram. On the other hand, their incommensurate phase (III) fits
roughly within the region of E-II phase of Plekhanov.

\begin{figure}[h]
\begin{center}
\includegraphics[scale=0.6]{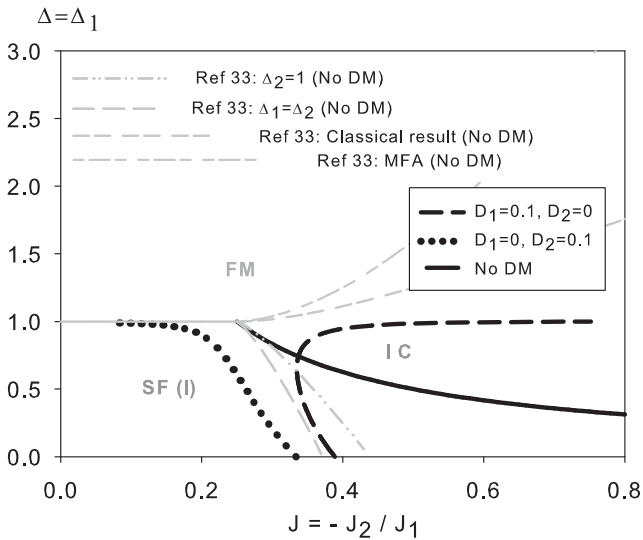}
\end{center}
\caption{The phase diagram according to ref. 35. Our results are shown in the
solid line (no DM interaction), the dashed line ($D_{1}=0.1,D_{2}=0$), and the
dotted line ($D_{1}=0,D_{2}=0.1$). SF (I) and IC denote spin-fluid (I) and
incommensurate phases. }%
\label{compareprb}%
\end{figure}

Since all the above calculation did not consider DM interaction, we set
$D_{1}=D_{2}=0$ at beginning in order to make comparison. This, in our
calculation, is to set $\Delta_{1}=\cos(\phi^{\prime}),\Delta_{2}=\cos
(2\phi^{\prime})$. The exact solution can be found when Eq. (\ref{nodm-eq-f})
is satisfied. We consider the case $J_{2}>0$. The notations of the phase
diagram of Dmitriev and Krivnov \cite{Dv07} was chosen because their mean
field calculation gives clear boundaries between phases. Besides, the result
of Plekhanov et al.\cite{Ple10} is consistent with that of Dmitriev and
Krivnov \cite{Dv07}. Our result is shown by the heavy solid line in Fig.
\ref{compareprb}. The ground states are the spiral spin states with different
pitch angles. The exact solution is in the region of incommensurate phase or
the E-II phase of ref. 33 or that of spin liquid III of ref. 35. Our
calculation shows that on the solid line the one-magnon state is gapless and
the chiral correlation has long-range order behavior due to spiral spins. For
larger $\Delta_{1}$, numerical calculations, except for ref. 33, found a
gapless chiral phase. For smaller $\Delta_{1}$, it has been argued
\cite{Jaf07} that it is a phase (denoted by spin fluid I) of XXZ model without
long-range order. Our calculation also shows that one-magnon modes actually
have lower energy than that of spiral spin state as it can be seen from Eq.
(\ref{symmetry-wk}). Hence, it is plausible to assume that the solid curve is
the \emph{exact} boundary of a gapless chiral phase. We noticed that our curve
is close to those given in ref. 35 in the region $0.25\leq J=-J_{2}/J_{1}%
\leq0.28$ but the difference increases as $J_{2}/J_{1}$ increases. Since the
calculation of Dmitriev and Krivnov \cite{Dv07} is less accurate for larger
$J_{2}/J_{1}$ whereas ours is exact, this discrepancy is expected. Further
study is needed to confirm that the curve by our calculation is indeed the
phase boundary. A final note of this case is that there exists another
\textquotedblleft mirror" curve if one makes the transformation: $\Delta
_{1}\rightarrow-\Delta_{1},J_{1}\rightarrow-J_{1},\phi^{\prime}\rightarrow
\pi-\phi^{\prime}$. This is actually equivalent to rotating the spins of odd
sites around \emph{z}-axis by $\pi$.

Next, we discuss the effect of DM interaction. With DM interaction, the
general form of the ratio of the couplings changes from Eq. (\ref{nodm-eq-f})
to the following
\begin{equation}
\frac{J_{1}}{J_{2}}=-2\frac{\sin(\phi_{2}-2\phi^{\prime})}{\sin(\phi_{1}%
-\phi^{\prime})}\frac{\cos(\phi_{1})}{\cos(\phi_{2})}, \label{eq-f}%
\end{equation}
as was given in Eq. (\ref{j1j2}). The system is more apt to have a spiral spin
ground state. For example, in Fig. \ref{compareprb}, the heavy dashed ($D_{1}
=0.1$) and dotted lines ($D_{2} =0.1$) indicate the conditions of such ground
states. In these calculations there is a small but non-vanishing $\Delta_{2}$.
Fortunately the result of Dmitriev and Krivnov \cite{Dv07} shows that the
effect of $\Delta_{2}$ will not influence the phase diagram very much. The
curves evidence that the spiral spin states can be the ground state in the
spin-fluid I phase of ref. 33. In other words, DM interaction makes chiral
correlation more robust.

For the case $J_{1} >0$ which was considered by Hikihara et al. \cite{Hik01},
if there is no DM interaction, there is no spiral spin ground state. The
reason is simple. In view of Eq. (\ref{nodm-eq-f}) $\cos(\phi^{\prime})$ has
to be negative for a positive $J_{1} / J_{2}$. This implies that $\Delta_{1} =
\cos(\phi^{\prime}) < 0$, is contradict to our starting point of $J_{1} >0$.
But if the DM interaction is present, the spiral spin states can still be the
ground state as long as the condition Eq. (\ref{eq-f}) is satisfied. For
example, we can choose $\phi^{\prime0}$. The region of PSD is shown in
Fig.\ref{example-psd-region}. We may have many possibilities. Some of them
are
\begin{align*}
&  (1.244,3.50,0,3.73,0.176),\phi_{1}=75^{0}\mbox{ and }\phi_{2}=10^{0}.\\
&  (5.85,1.97,0,1.73,0.176),\phi_{1}=60^{0}\mbox{ and }\phi_{2}=10^{0}.\\
&  (1.28,3.50,0.185,3.73,0.364), \phi_{1}=75^{0}\mbox{ and }\phi_{2}=20^{0}%
\end{align*}
where the parentheses give the values of $(J_{1} / J_{2},\Delta_{1},\Delta
_{2},D_{1},D_{2})$. Under these conditions, we have $J_{1}>0,J_{2}>0$. Hence,
due to the effect of DM interaction, the spiral spin states can be the ground
state in the regions of the spin-fluid phase and dimer phase designated by
Hikihara et al. \cite{Hik01} even if the magnitude of the DM interaction
should not be small.

There have been many numerical calculations on the zigzag spin chain system.
However, due to the large number of parameters, and hence, large dimension of
phase space, and the different methods of calculation, it is not always clear
how physical properties change due to the interplays among these parameters.
For example, how the system evolves if $\Delta_{1} \Delta_{2} <0$ but its
magnitude increases? A very helpful way is to calculate the correlation
functions as it has been done in many works. In many cases, finding exact
solutions, albeit in a very limited region can provide more clues. A simple
calculation shows that the spiral spin state is able to sustain not only the
vector chiral spin order but also the oscillatory spin correlation on the
easy-plane
\[
\langle\sum_{j} s_{j}^{x}s_{j+d}^{x}\rangle=\frac{N}{8}\sin^{2}(\beta
)\cos(\phi^{\prime}),
\]
where $\phi^{\prime}$ is the spiral angle and $\beta$ is the cone angle. The
correlation should oscillate but not decay. Numerical analysis should be able
to find long-range order behavior of the chiral correlation function and the
$xy$-plane oscillatory spin correlation function. But there is some difficulty
due to the symmetry of the Hilbert space caused by the degenerate ground
states as we have discussed in section IV. As it has been mentioned, if the
anisotropic exchange interaction can be transformed away then it can be shown
that all the FP states are the ground states of the transformed Hamiltonian.
This implies that our Hamiltonian has huge degeneracy. The ground state can be
any linear combination of these degenerate states. If in numerical
calculation, the sample of ground states is not large enough, the symmetry can
be lost and correlation functions decay faster than they should be.

\section{Conclusions \label{summary}}
To find the conditions for spiral spin state as the ground state, we transform
the physical Hamiltonian to a Hamiltonian with isotropic exchange interaction
and DM interaction. The FP states are the eigen states of the transformed
Hamiltonian. Then we used positive semi-definite theorem to identify the
region of FP state being the GS for the transformed Hamiltonian, which is
nothing but the same region of spiral spin state as GS of the original
Hamiltonian. The region (shown in Fig. \ref{psd-region-p1-p2}) can be
expressed by a very simple relation with the couplings of NN and NNN
superexchange interaction and DM interaction. The effect of DM interaction is
important because its strength is related to the pitch angle of spiral spins
and the unitary transformation. As the strength of the DM interaction
increases, the pitch angle $\phi^{\prime}$ also increases. It was also found
that for spiral spin states to be GS, either $2\phi_{1}=2\tan^{-1}(D_{1}%
)\geq2\phi_{2}=2\tan^{-1}(D_{2})$ and $\phi_{1}\leq\phi^{\prime}\leq\phi
_{2}/2+\pi/4$, or $\phi_{2}/2-\pi/4\leq\phi^{\prime}\leq\phi_{2}/2$ and
$\phi^{\prime}\leq\phi_{1}\leq\phi_{2}/2+\pi/2$. Wherever the equal signs
stand, the spiral spin states are degenerate with one-magnon states for the
physical Hamiltonian in Eq. (\ref{eq-H}). Hence, the boundary in Fig.
\ref{psd-region-p1-p2} marks the region where the spiral spin states being the
GS. We thus made manifestly the relation between DM interaction and spiral
spin states.

These results can serve as a guided line for experimentalists to find spiral
spin state which, in turn, can lead to multiferroics. Our results also show
the connection between spiral spins and magnetic frustration. Finally, our
method can be applied to other types of magnetic system such as coupled spin
chains and layer structure. It is not restricted to spin-$1/2$ system, but can
be applied to any spin system.

\begin{acknowledgments}
This work is supported by the National Science Council of R.O.C. under grant
number NSC 98-2112-N-002-001-MY3.
\end{acknowledgments}


\end{document}